\documentclass[runningheads]{llncs}
\usepackage[T1]{fontenc}
\usepackage{graphicx}
\usepackage{fontspec}   
\usepackage{amsmath, amssymb, amsfonts}
\usepackage{booktabs}    
\usepackage{multirow}    

\usepackage{marvosym} 

\begin{document}
\title{PGSTalker: Real-Time Audio-Driven Talking Head Generation via 3D Gaussian Splatting with Pixel-Aware Density Control}
\titlerunning{PGSTalker}
%




%

\author{
  \small Tianheng Zhu\inst{1} \and
   \small  Yinfeng Yu\inst{1}\textsuperscript{(\Letter)} \and
    \small Liejun Wang\inst{1} \and
    \small Fuchun Sun\inst{2} \and
    \small Wendong Zheng\inst{3}
}

\authorrunning{T. Zhu et al.}

\institute{
   \small  Xinjiang Multimodal Intelligent Processing and Information Security Engineering Technology Research Center, \\
    \small School of Computer Science and Technology, Xinjiang University, Urumqi 830017, China \\
  \email{yuyinfeng@xju.edu.cn} \and
   \small  Department of Computer Science and Technology, Tsinghua University, Beijing 100091, China \and
   \small  School of Electrical Engineering and Automation, Tianjin University of Technology, Tianjin 300382, China
}

\renewcommand{\thefootnote}{}  
\footnotetext{\textsuperscript{(\Letter)} \small Yinfeng Yu is the corresponding author (e-mail: yuyinfeng@xju.edu.cn).}

\maketitle              

\begin{figure}
    \centering
	\includegraphics[width=\textwidth]{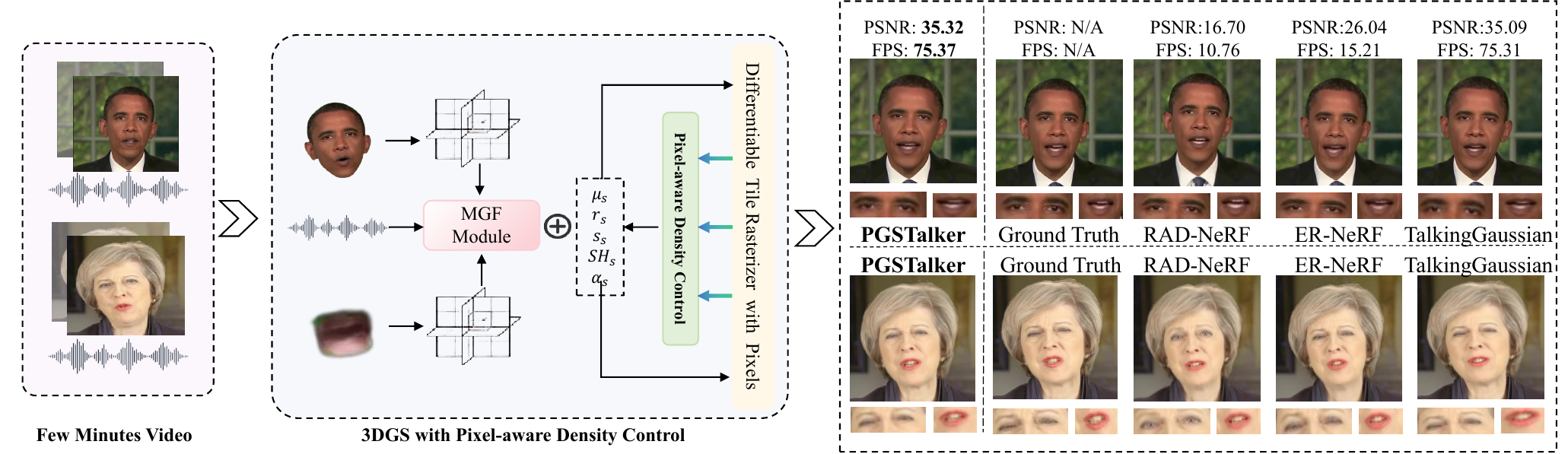}
	\caption{Our PGSTalker introduces a pixel-aware density control strategy that adaptively guides point cloud splitting, enhancing detail in dynamic areas. This enables a faster, more accurate, and high-quality talking head synthesis approach.} 
	\label{headfig}
\end{figure}

\begin{abstract}
Audio-driven talking head generation is crucial for applications in virtual reality, digital avatars, and film production. While NeRF-based methods enable high-fidelity reconstruction, they suffer from low rendering efficiency and suboptimal audio-visual synchronization. This work presents PGSTalker, a real-time audio-driven talking head synthesis framework based on 3D Gaussian Splatting (3DGS). To improve rendering performance, we propose a pixel-aware density control strategy that adaptively allocates point density, enhancing detail in dynamic facial regions while reducing redundancy elsewhere. Additionally, we introduce a lightweight Multimodal Gated Fusion Module to effectively fuse audio and spatial features, thereby improving the accuracy of Gaussian deformation prediction. Extensive experiments on public datasets demonstrate that PGSTalker outperforms existing NeRF- and 3DGS-based approaches in rendering quality, lip-sync precision, and inference speed. Our method exhibits strong generalization capabilities and practical potential for real-world deployment.

\end{abstract}

\section{Introduction}

Audio-driven talking head generation has received increasing attention in computer vision and graphics, with applications in digital humans, virtual reality, teleconferencing, and film production~\cite{introduce1,in2,wav2lip,in3,synobama,nvp,x2face,sadtalker}. Despite notable progress, existing methods still face challenges in rendering efficiency and audio-visual synchronization.

Neural Radiance Field (NeRF)-based approaches~\cite{nerf,guo2021ad,tang2022real,ye2023geneface} leverage implicit 3D modeling and audio-conditioned MLPs to synthesize facial motion, but often suffer from limited lip-sync accuracy, high computational cost, and slow rendering speed. Even with recent advances such as hash encoding and attention mechanisms~\cite{ernerf,synctalk}, their implicit nature constrains real-time performance.

Alternatively, 3D Gaussian Splatting (3DGS)~\cite{3dgs} enables efficient explicit rendering using Gaussian point clouds, supporting fast and high-quality scene synthesis. Recent works have explored its application to talking head generation~\cite{4dgs,dynamic3dgs,deformable3dgs,photo3dgs,li2025talkinggaussian,cho2024gaussiantalker}. 
Still, two key challenges remain: (1) generation quality is susceptible to point cloud initialization, where Structure-from-Motion (SfM)-based methods often introduce artifacts in fine regions (e.g., teeth); and (2) high-density initialization using 3DMMs enhances detail but significantly reduces rendering speed, limiting real-time applicability.

To address these limitations, we propose \textbf{PGSTalker}, a real-time audio-driven talking head generation framework based on 3DGS. We introduce a pixel-aware density control strategy that refines the point cloud based on per-Gaussian pixel coverage, assigning higher density to dynamic regions (e.g., lips, eyes) while maintaining sparsity in static areas. This improves rendering precision without sacrificing speed. Additionally, we design a lightweight Multimodal Gated Fusion (MGF) module to adaptively fuse audio and spatial features for accurate Gaussian deformation prediction. Unlike prior MLP-based or attention-based designs~\cite{li2025talkinggaussian,cho2024gaussiantalker}, our module dynamically modulates feature interaction, enhancing deformation accuracy with minimal overhead and strong real-time performance.

The main contributions of this paper are as follows:

\begin{itemize}
    \renewcommand{\labelitemi}{\Large$\bullet$}
    \item We propose a novel 3DGS-based framework for talking head generation. We introduce a pixel-aware density control strategy that adaptively splits Gaussians based on their pixel contributions, enabling efficient rendering with enhanced spatial detail in dynamic regions.
    
    \item We design a new multimodal fusion mechanism that adaptively learns weights between audio and spatial features, significantly improving the accuracy and efficiency of audio-driven Gaussian deformation prediction.
    
    \item Extensive experiments demonstrate that PGSTalker achieves superior performance in rendering quality, lip-sync accuracy, and inference speed, with strong generalization and practical application potential.
\end{itemize}

\section{Related Work}

Audio-driven talking head synthesis aims to generate temporally coherent and realistic facial videos from speech. Early 2D methods~\cite{introduce1,in2,tvsa,wav2lip,x2face} relied on image-based generation but lacked robust 3D modeling, resulting in limited pose controllability and temporal consistency. Subsequent works~\cite{lsp,nvp,mead,facial,sadtalker} introduced intermediate representations (e.g., landmarks, meshes) to enhance motion modeling, though they often remained sensitive to estimation noise. 

Neural Radiance Field (NeRF)-based approaches~\cite{nerf,guo2021ad,tang2022real,ernerf} enabled high-fidelity 3D synthesis by conditioning radiance fields on audio but incurred high computational costs. Recent methods~\cite{ye2023geneface,geneface++,synctalk} improved generalization using pretrained audio encoders; however, instability in dynamic regions persists due to color field modulation.

Meanwhile, 3D Gaussian Splatting (3DGS)~\cite{3dgs} has emerged as an efficient alternative, supporting real-time rendering through explicit Gaussian primitives. TalkingGaussian~\cite{li2025talkinggaussian} explored lightweight deformation and dual-branch fusion but lacked expressive multimodal modeling. GaussianTalker~\cite{cho2024gaussiantalker} initializes the point cloud using a 3D Morphable Model (3DMM) and introduces cross-attention to enhance generation quality, but suffers from high computational overhead and point redundancy.

In contrast, our proposed \textbf{PGSTalker} framework employs a pixel-aware density control strategy to balance rendering precision and efficiency. It integrates a multimodal gated fusion module to capture complex audio-spatial interactions, enabling accurate Gaussian deformation and real-time, high-fidelity talking head synthesis.

\section{Methods}
\subsection{Preliminary}

\subsubsection{3D Gaussian Splatting.}

3D Gaussian Splatting (3DGS)~\cite{3dgs} is an explicit 3D scene representation using point-based Gaussian primitives. Each 3D Gaussian is parameterized by a set of optimizable variables, including the mean position $\mu$, a positive semi-definite covariance matrix $\Sigma$, spherical harmonics coefficients $C$, and opacity $\alpha$, formulated as $G(x)=e^{-\frac{1}{2}(x-\mu)^T\Sigma^{-1}(x-\mu)}$.

To enable differentiable optimization, the covariance matrix $\Sigma$ is typically decomposed into a scaling matrix $S$ and a rotation matrix $R$, such that $\Sigma = R S S^T R^T$. These matrices are further parameterized by two vectors: a 3D scaling vector $s \in \mathbb{R}^3$ and a quaternion rotation vector $r \in \mathbb{R}^4$.

During rendering, each 3D Gaussian is projected onto the image plane using splatting techniques. As described in~\cite{zwicker2001ewa}, given a view transformation $S$ and a Jacobian matrix $J$ from a projective transformation, the transformed covariance matrix $\Sigma^{\prime}$ in ray space is computed as $\Sigma^{\prime} = J W \Sigma W^T J^T$.

The color $C$ and opacity $\alpha$ at each image pixel are then computed by blending overlapping Gaussians using weighted accumulation: $C = \sum_{i \in N} c_i \alpha_i \prod_{j=1}^{i-1} (1 - \alpha_j)$, where $c_i$ and $\alpha_i$ denote the color and opacity of the $i$-th Gaussian. This compositing process ensures that each pixel color is determined by the contribution of all overlapping Gaussians.

\subsubsection{Adaptive Density Control of 3DGS.}

To achieve adaptive point cloud refinement in the 3DGS framework~\cite{3dgs}, a gradient-based splitting or cloning strategy is employed to progressively optimize the spatial distribution of Gaussians. This mechanism is typically triggered at fixed training intervals. Specifically, if the screen-space gradient magnitude of a Gaussian exceeds a threshold $\tau_{\text{pos}}$ across multiple views, the point is considered important and is duplicated to improve representation.

Formally, for a Gaussian point $i$ observed in $M^i$ views, the average gradient in the Normalized Device Coordinates (NDC) space is given by:

\begin{equation}
	\frac{1}{M^i} \sum_{k=1}^{M^i} \sqrt{
		\left(\frac{\partial \mathcal{L}_k}{\partial \mu^{i,k}_{\text{ndc},x}}\right)^2 +
		\left(\frac{\partial \mathcal{L}_k}{\partial \mu^{i,k}_{\text{ndc},y}}\right)^2}
	> \tau_{\text{pos}},
	\label{eq:adaptive_density}
\end{equation}

Here, $\mu^{i,k}_{\text{ndc},x}$ and $\mu^{i,k}_{\text{ndc},y}$ are the screen-space coordinates of Gaussian $i$ in view $k$, and $\mathcal{L}_k$ is the rendering loss for that view. If the condition is satisfied, the point undergoes a split or clone operation to refine the density in critical regions.

\begin{figure}
	\includegraphics[width=\textwidth]{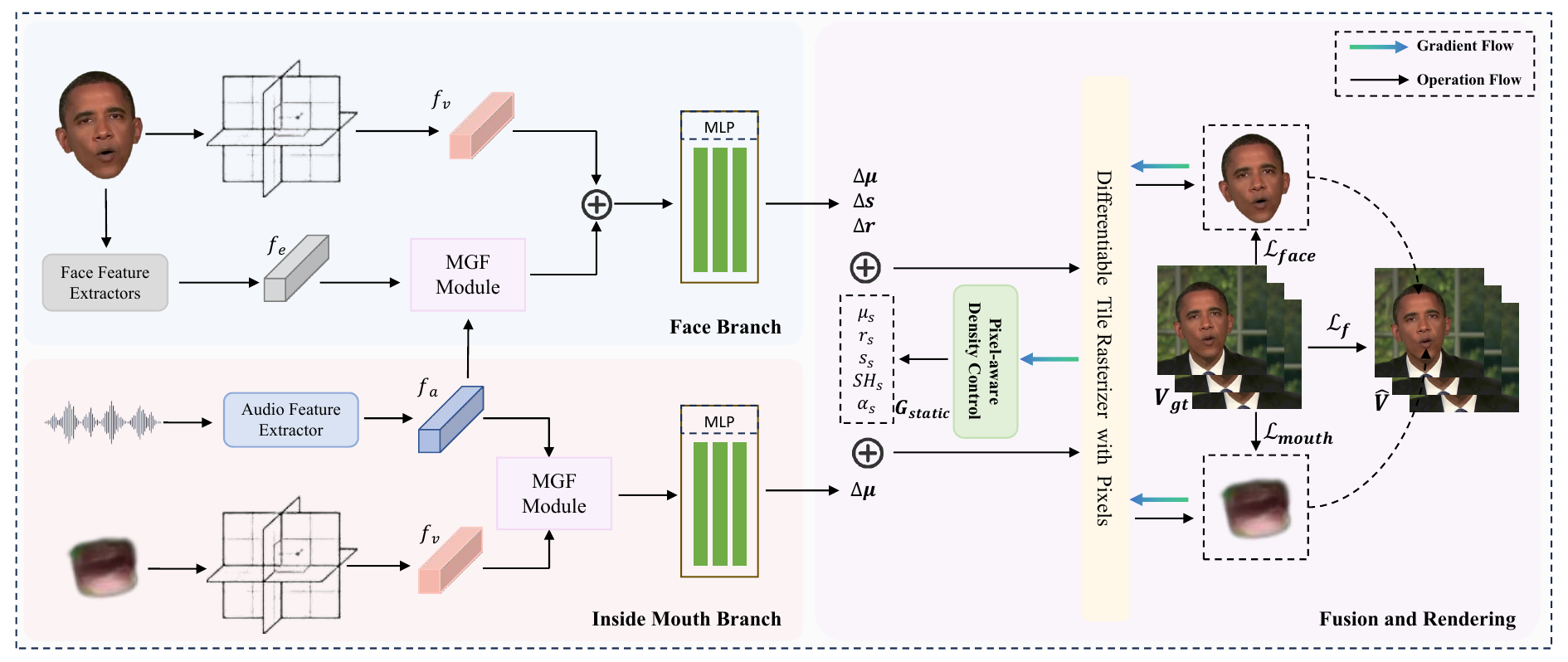}
	\caption{Overview of the PGSTalker framework. It consists of a Face Branch and an Inside Mouth Branch, which model different facial regions. Audio and spatial features are fused via a Multimodal Gated Fusion (MGF) module to predict Gaussian deformations. The output is rendered using pixel-aware density control and composited into high-quality talking head images.} 
	\label{framework}
\end{figure}

\subsection{Overview}

As illustrated in Fig.~\ref{framework}, PGSTalker learns from an audio-video sequence containing a set of training frames $\mathcal{I}=\left\{I_0, I_1, ..., I_n\right\}$ by constructing two dedicated branches: a Face Branch and an Inside Mouth Branch, each modeling distinct facial motion regions. For both branches, we initialize a 3D Gaussian point cloud $\theta_{C}=\{\mu_c, s_c, q_c, \alpha_c, SH_c\}$, where $\mu_c$ denotes position, $s_c$ scaling, $q_c$ rotation, $\alpha_c$ opacity, and $SH_c$ spherical harmonics for color.

We begin by generating a randomly initialized point cloud from static frames and optimizing it using averaged scene priors and grid-based motion fields to obtain a structured 3D representation. However, this initialization lacks explicit spatial encoding and struggles to capture complex facial dynamics. To enhance spatial expressiveness, we adopt a tri-plane hash position encoder $\mathcal{H}$~\cite{li2025talkinggaussian} to extract spatial features $f_v$ for each point. These features are fused with audio features $f_a$ and expression features $f_e$ to predict deformation parameters $\delta_{i} = \{\Delta \mu_i, \Delta s_i, \Delta q_i\}$, which drive audio-based dynamic motion. Notably, only position, scale, and rotation are updated in this stage, while color and opacity remain fixed.

To facilitate effective feature fusion, we propose a Multimodal Gated Fusion Module (MGF) that adaptively modulates interactions between audio and spatial cues, enabling more accurate and efficient deformation prediction (Section~\ref{sec:mgf}). The deformed point clouds are then rendered using a differentiable tile-based rasterizer with pixel-aware parameters. Pixel-wise gradients are backpropagated to the density control module to refine the 3D structure, as detailed in Section~\ref{sec:pixelaware}. Finally, the rendered outputs from both branches are composited to generate high-fidelity talking head images.

\subsection{Pixel-aware Density Control}
\label{sec:pixelaware}

Traditional 3DGS~\cite{3dgs} adopts a gradient-based density control strategy, where a Gaussian point is split or cloned if the average screen-space gradient of its NDC projection across multiple views exceeds a threshold $\tau_{\mathrm{pos}}$. However, this uniform treatment is suboptimal for talking head synthesis, particularly in highly dynamic regions (e.g., mouth, eyebrows) and along structural boundaries (e.g., facial contours). It fails to capture the actual contribution of each point to rendered pixels, often suppressing point growth in critical areas.

To address this, we introduce a pixel-aware density control strategy inspired by Pixel-GS~\cite{pixelgs}, where the gradient is weighted by the number of pixels covered by each Gaussian across views. For each point $i$ observed in $M^i$ views, the pixel-aware gradient is defined as:

\begin{equation}
	\frac{\sum_{k=1}^{M^i} m^i_k \cdot \left\| \nabla_{\mu^{i,k}_{\mathrm{ndc}}} \mathcal{L}_k \right\|}{
		\sum_{k=1}^{M^i} m^i_k} > \tau_{\mathrm{pos}},
	\label{eq:pixel_aware_gradient}
\end{equation}

where $m^i_k$ denotes the number of pixels covered by point $i$ in view $k$, and $\left\| \nabla_{\mu^{i,k}_{\mathrm{ndc}}} \mathcal{L}_k \right\|$ is the 2D screen-space gradient magnitude:

\begin{equation}
\left\| \nabla_{\mu^{i,k}_{\mathrm{ndc}}} \mathcal{L}_k \right\| =
\sqrt{
	\left(\frac{\partial \mathcal{L}_k}{\partial \mu^{i,k}_{\mathrm{ndc},x}}\right)^2 +
	\left(\frac{\partial \mathcal{L}_k}{\partial \mu^{i,k}_{\mathrm{ndc},y}}\right)^2
}.
\end{equation}

Following Zhang et al., we improve robustness by filtering out Gaussians that are either too close to the camera or outside image bounds. A point $i$ in view $k$ is considered valid only if its camera-space depth $\mu^{i,k}_{c,z} > 0.2$ and its projected center $(\mu^{i,k}_{p,x}, \mu^{i,k}_{p,y})$ with radius $R^k_i$ falls within the image:

\begin{equation}
	\begin{cases}
		R^i_k > 0,\quad \mu^{i,k}_{c,z} > 0.2, \\
		-R^i_k - 0.5 < \mu^{i,k}_{p,x} < R^i_k + W - 0.5, \\
		-R^i_k - 0.5 < \mu^{i,k}_{p,y} < R^i_k + H - 0.5.
	\end{cases}
\end{equation}

Here, $W$ and $H$ are the image width and height. We further refine $m^i_k$ by considering only pixels that contribute meaningfully to rendering. A pixel $(x, y)$ is counted if it lies within $R^k_i$ of the projected center, has sufficient opacity $\alpha^{i}_{k,(x,y)} \geq \frac{1}{255}$, and is not fully occluded by prior points, i.e., $\prod_{j=1}^{i-1}(1 - \alpha^j_{k,(x,y)}) \geq 10^{-4}$:

\begin{equation}
\sqrt{(x - \mu^{i,k}_{p,x})^2 + (y - \mu^{i,k}_{p,y})^2} < R^k_i.
\end{equation}

This pixel-aware weighting strategy preserves the sparsity control of standard 3DGS while significantly improving detail modeling in dynamic facial regions (e.g., lip motion, eyebrow raising) and along structure boundaries. It leads to better lip-sync accuracy and visual fidelity, especially under complex, fast-changing audio inputs. Importantly, the method introduces only lightweight pixel statistics, adding negligible computational overhead during training and rendering.

\begin{figure}
    \centering
	\includegraphics[width=0.95\textwidth]{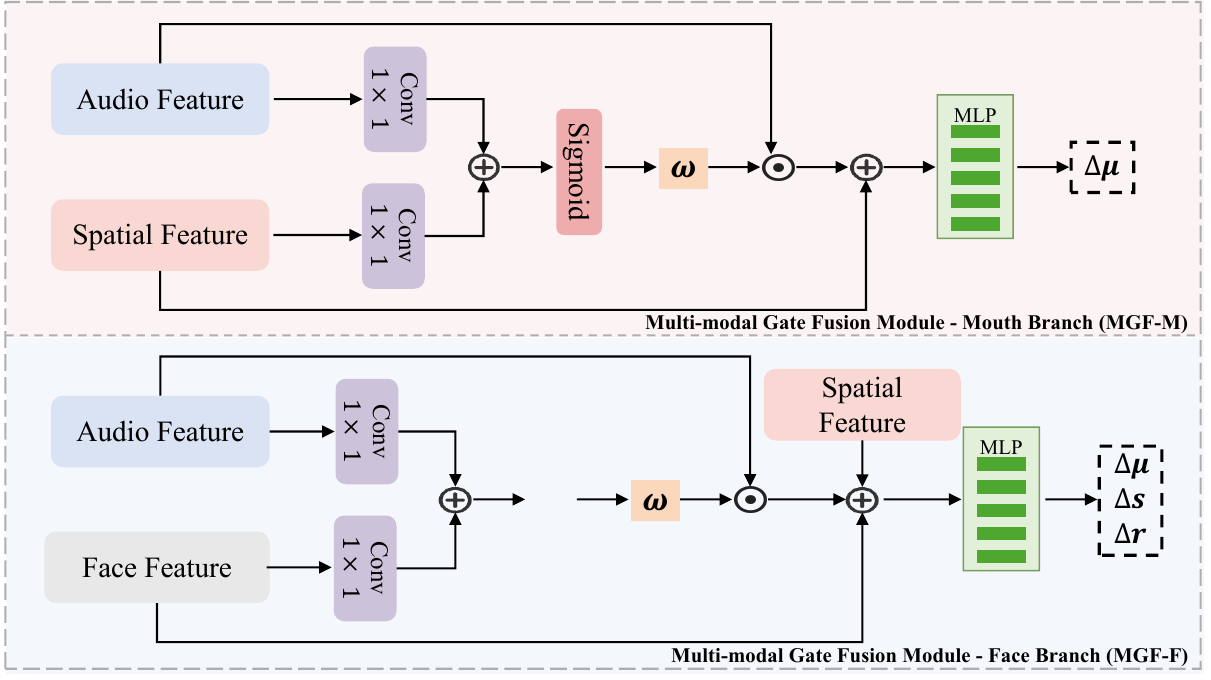}
	\caption{Architecture of the Multimodal Gated Fusion (MGF) module used for predicting Gaussian point deformations in both facial and inside-mouth regions. Audio and spatial/expression features are first processed via $1 \times 1$ convolutions, then concatenated to generate gating weights that modulate the audio features. The modulated audio features are fused with spatial residuals and passed through an MLP to predict deformation parameters. This design enables adaptive and effective fusion of multimodal cues for precise and expressive Gaussian deformation.}
	\label{mgf}
\end{figure}

\subsection{Audio-Driven Gaussian Deformation Prediction}
\label{sec:mgf}

Static 3D Gaussian points must deform in response to speech input to achieve high-fidelity, audio-synchronized talking head generation. While prior methods~\cite{li2025talkinggaussian} typically use audio features as direct input to MLPs for predicting Gaussian deformations, they often fail to capture the complex interactions between audio and spatial features, leading to unnatural motion and poor lip synchronization. Other approaches incorporate attention mechanisms~\cite{cho2024gaussiantalker}, but incur high computational overhead.

To overcome these limitations, we propose a lightweight yet expressive Multimodal Gated Fusion (MGF) module (Fig.~\ref{mgf}) to fuse audio with spatial or expression features more effectively. Separate MGF modules are employed for the face and inside-mouth regions to account for their distinct motion patterns.

\subsubsection{Inside Mouth Region}
\label{sec:mouthfusion}

Since mouth movement is primarily audio-driven, we employ an audio-centric fusion scheme. Spatial features $f_s$ and audio features $f_a$ are first projected via $1 \times 1$ convolutions into $f_s'$ and $f_a'$, concatenated, and passed through a Sigmoid-activated convolution to produce a gating vector $\omega$:

\begin{equation}
	\omega = \sigma\left(\text{Conv}_{1\times1}\left([f_s'; f_a']\right)\right), \quad \tilde{f}_a = \omega \odot f_a'.
\end{equation}

The modulated audio feature $\tilde{f}_a$ is concatenated with $f_s'$ to obtain the fused feature $f_{fuse}^{mouth} = \text{Concat}(f_s', \tilde{f}_a)$, which is fed into an MLP to predict position deformation $\Delta \mu_i$:

\begin{equation}
	\Delta \mu_i = \text{MLP}(f_{fuse}^{mouth}).
\end{equation}

\subsubsection{Face Region}
\label{sec:facefusion}

Facial movements such as eye blinking or brow raising are driven by audio and non-verbal cues. To capture these dependencies, we fuse audio features $f_a$ with facial expression features $f_e$ using a similar gated mechanism:

\begin{equation}
	\omega = \sigma\left(\text{Conv}_{1\times1}\left([f_a'; f_e']\right)\right), \quad \tilde{f}_a = \omega \odot f_a', \quad f_{ae} = \text{Concat}(\tilde{f}_a, f_e').
\end{equation}

To avoid interference with mouth modeling, we follow~\cite{li2025talkinggaussian} and use only non-mouth Action Units~\cite{eye} for $f_e$. The final fused representation is obtained as $f_{fuse}^{face} = \text{Concat}(f_a, f_{ae}, f_s)$, which is input to an MLP to predict full deformation parameters:

\begin{equation}
	\{\Delta \mu_i, \Delta s_i, \Delta q_i\} = \text{MLP}(f_{fuse}^{face}).
\end{equation}

\subsubsection{Fusion and Rendering}

The deformed point clouds from the face and mouth branches are rendered independently. The final image is composited by overlaying the mouth behind the face region based on opacity $\mathcal{A}_{\text{face}}$:

\begin{equation}
	\mathcal{C}_{\text{head}}(\mathbf{x}_p) = \mathcal{C}_{\text{face}}(\mathbf{x}_p) \cdot \mathcal{A}_{\text{face}}(\mathbf{x}_p) + \mathcal{C}_{\text{mouth}}(\mathbf{x}_p) \cdot (1 - \mathcal{A}_{\text{face}}(\mathbf{x}_p)).
\end{equation}

\vspace{-0.5em}
\subsection{Training Objective}

We adopt a three-stage training pipeline inspired by~\cite{li2025talkinggaussian}: static initialization, deformation prediction, and final fine-tuning. Each branch (face and mouth) is trained independently in the first two stages and jointly optimized in the final stage.

\subsubsection{Static Initialization}

We optimize initial Gaussian parameters $\theta_C = \{\mu, s, q, \alpha, f\}$ using masked supervision on static frames. The loss combines L1 and D-SSIM:

\begin{equation}
	\mathcal{L}_{C} = \mathcal{L}_{1}(\hat{\mathcal{I}}_{C}, \mathcal{I}_{\text{mask}}) + \lambda \cdot \mathcal{L}_{\text{D-SSIM}}(\hat{\mathcal{I}}_{C}, \mathcal{I}_{\text{mask}}).
\end{equation}

\subsubsection{Gaussian Deformation Prediction}

We train the network to predict audio-driven deformations $\delta = \{\Delta \mu, \Delta s, \Delta q\}$ and update dynamic parameters $\theta_D = \{\mu + \Delta \mu, s + \Delta s, q + \Delta q, \alpha, SH\}$. The supervision follows the same loss formulation:

\begin{equation}
	\mathcal{L}_{D} = \mathcal{L}_{1}(\hat{\mathcal{I}}_{D}, \mathcal{I}_{\text{mask}}) + \lambda \cdot \mathcal{L}_{\text{D-SSIM}}(\hat{\mathcal{I}}_{D}, \mathcal{I}_{\text{mask}}).
\end{equation}

\subsubsection{Fine-tuning}

In the final stage, we refine the rendered head image $\hat{\mathcal{I}}_{\text{head}}$ against the full ground truth $\mathcal{I}_{gt}$ using a combination of pixel-wise and perceptual losses:

\begin{equation}
	\mathcal{L}_{F} = \mathcal{L}_{1}(\hat{\mathcal{I}}_{\text{head}}, \mathcal{I}_{gt}) + \lambda \cdot \mathcal{L}_{\text{D-SSIM}}(\hat{\mathcal{I}}_{\text{head}}, \mathcal{I}_{gt}) + \gamma \cdot \mathcal{L}_{\text{LPIPS}}(\hat{\mathcal{I}}_{\text{head}}, \mathcal{I}_{gt}).
\end{equation}

Only the color-related parameters $SH \in \theta_C$ are updated in this stage, while point density remains fixed for stability. This hierarchical training strategy enables the model to learn structure, motion, and appearance progressively for realistic and synchronized talking head generation.

\section{Experiment}
\subsection{Experimental Settings}

\begin{table}
	\caption{The quantitative results of the \textbf{self-driven} setting. The best and second-best methods are in \textbf{bold} and \underline{underline}, respectively. Our PGSTalker achieves competitive results while significantly improving inference efficiency.
	}
	\label{tab:db}
	\centering
	\begin{tabular}{ccccccccc}
		\hline
		\multirow{2}{*}{\textbf{Methods}} & \multicolumn{3}{c}{\textbf{Rendering Quality}} & \multicolumn{3}{c}{\textbf{Motion Quality}} & \multicolumn{2}{c}{\textbf{Efficiency}}          \\  
		& PSNR$\uparrow$     & SSIM$\uparrow$     & LPIPS$\downarrow$       & LMD$\downarrow$     & LSE-D$\downarrow$   & LSE-C$\uparrow$               & Time$\downarrow$       & \multicolumn{1}{c}{FPS$\uparrow$}          \\ \hline
		GT & N/A &1   &0     &0   & 6.745   & 8.491      & N/A & \multicolumn{1}{c}{N/A} \\
		Wav2Lip                           & 16.507   & 0.847    & 0.2994        & 4.978   & \textbf{7.019}   & \textbf{7.972}     & -   & \multicolumn{1}{c}{8.759}        \\
		Ip-LAP                            & 16.418   & 0.841    & 0.2992         & 4.665   & 8.99    & 5.256    & -            & \multicolumn{1}{c}{1.314}        \\
		AD-NeRF                           & 25.794   & 0.9643   & 0.0842          & 2.932   & 9.839   & 5.105      & 167.7h      & \multicolumn{1}{c}{0.04}         \\
		RAD-NeRF                          & 16.706   & 0.855    & 0.281           & 3.291   & 8.669   & 6.288      & 13.8h       & \multicolumn{1}{c}{10.76}        \\
		ER-NeRF                           & 26.047   & 0.961    & 0.0635         & 2.547   & 7.913   & 7.054      & 8.9h        & \multicolumn{1}{c}{15.21}        \\
		GaussianTalker                    & \textbf{36.034}   &\textbf{0.992}    & 0.0224           & 2.614   & 8.274   & 6.964      & 4.5h        & \multicolumn{1}{c}{61.24}        \\
		TalkingGaussian                    & 35.09   & 0.9901    & \underline{0.0197}           & \underline{2.532}   & 8.013   & 7.001      & \textbf{1.2h}        & \multicolumn{1}{c}{\underline{75.31}}        \\
		Ours(PGStalker)           & \underline{35.32}  & \underline{0.9903} & \textbf{0.0189}     & \textbf{2.469}   & \underline{7.855}   & \underline{7.266}      & \underline{1.5h}        & \multicolumn{1}{c}{\textbf{75.37}} \\ \hline
	\end{tabular}
\end{table}

\subsubsection{Dataset and Pre-processing}

We train a personalized model for each speaker using only a few minutes of video and corresponding audio. Specifically, we curate a dataset from four high-quality talking head videos~\cite{guo2021ad,ernerf,ye2023geneface}, featuring three male and one female speaker. Each video contains approximately 6500 frames at 25 FPS with a centered portrait-style talking face. Three videos are resized to $512 \times 512$, while the “Obama” video is $450 \times 450$. All experiments utilize a dataset partitioned into standard training and test splits. Computation is constrained to a single NVIDIA Tesla T4 GPU with 16GB of onboard memory.

\subsubsection{Comparison Baselines}

We evaluate PGSTalker against several state-of-the-art talking head generation methods. NeRF-based baselines include AD-NeRF~\cite{guo2021ad}, RAD-NeRF~\cite{tang2022real}, and ER-NeRF~\cite{ernerf}, which employ implicit 3D modeling. 3DGS-based methods include GaussianTalker~\cite{cho2024gaussiantalker} and TalkingGaussian~\cite{li2025talkinggaussian}, which model audio-driven deformation with explicit Gaussian point clouds. We also include two 2D-based approaches for broader comparison: Wav2Lip~\cite{wav2lip} and IP-LAP~\cite{iplap}.

\subsection{Quantitative Evaluation}

\subsubsection{Comparison Settings and Metrics}

Following prior works~\cite{guo2021ad,ernerf,cho2024gaussiantalker}, we evaluate performance under two settings: \textit{self-driven} and \textit{cross-driven}. In the self-driven setting, we assess identity-specific head reconstruction using the test set. Image quality is evaluated with standard metrics, including Peak Signal-to-Noise Ratio (\textbf{PSNR}), Structural Similarity Index (\textbf{SSIM})~\cite{SSIM}, and Learned Perceptual Image Patch Similarity (\textbf{LPIPS})~\cite{LPIPS}, all computed within the facial region.To measure audio-visual synchronization, we use Landmark Distance (\textbf{LMD}), Lip Sync Error – Distance (\textbf{LSE-D}), and Lip Sync Error – Confidence (\textbf{LSE-C})~\cite{wav2lip}. Additionally, training time (\textbf{Time}) and inference speed (\textbf{FPS}) are reported to compare overall efficiency.

In the cross-driven setting, models are tested with unrelated audio inputs to evaluate generalization in lip synchronization. Audio clips are sourced from the SynObama demo~\cite{synobama}. Since ground-truth frames are unavailable, only LMD, LSE-D, and LSE-C~\cite{wav2lip} are used for evaluation.

\begin{figure}
    \centering
	\includegraphics[width=\textwidth]{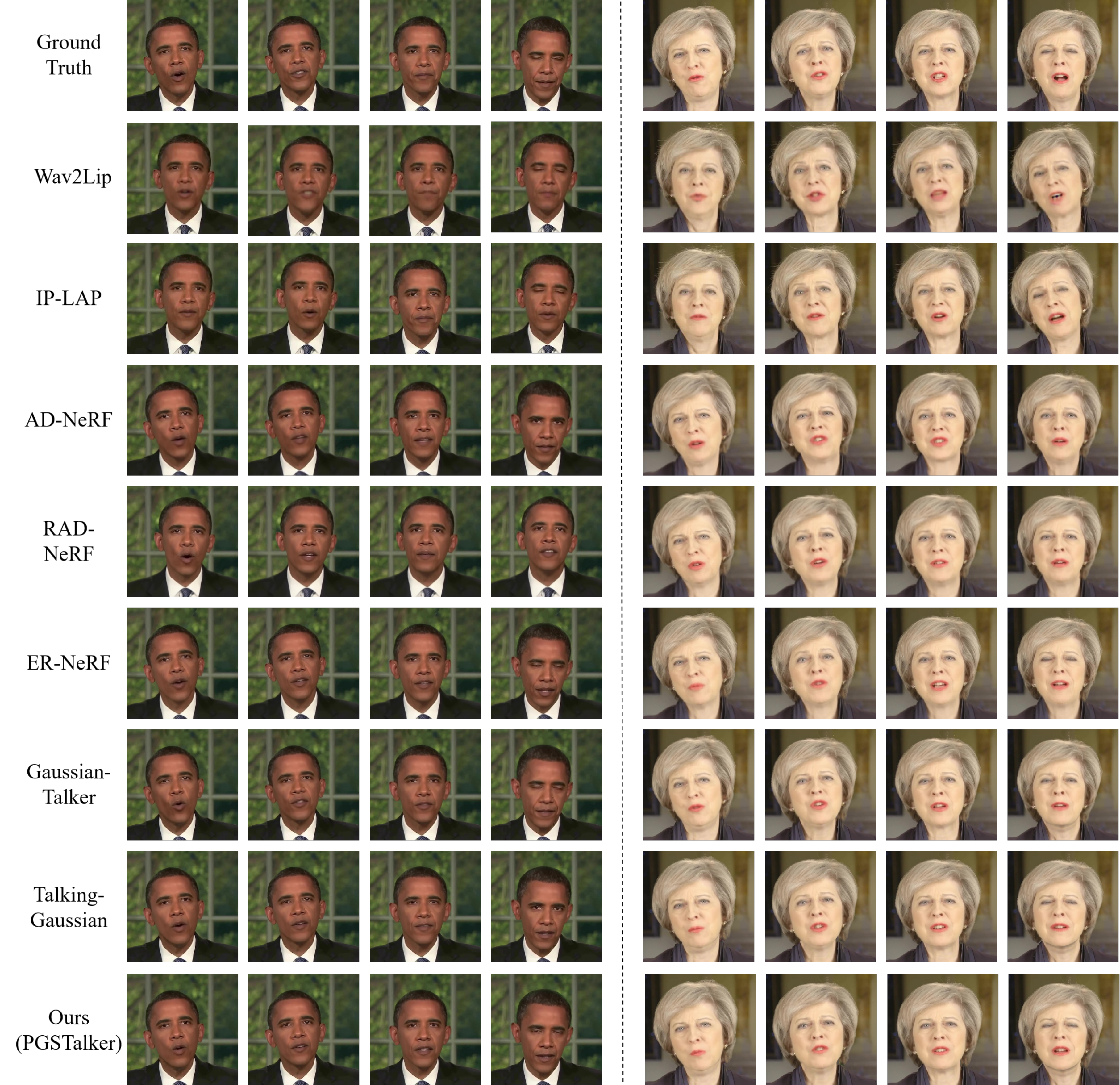}
	\caption{Qualitative results under the self-driven setting on the Obama and May datasets. Compared to advanced 3DGS-based talking head generation methods, our approach achieves competitive results in reconstruction quality and lip-sync accuracy, while providing better control over head pose and facial expressions.} 
	\label{result}
\end{figure}

\subsubsection{Self-driven Evaluation}

Table~\ref{tab:db} reports the quantitative results under the self-driven setting. PGStalker consistently outperforms other methods across rendering quality, audio-visual synchronization, and inference efficiency. Compared to 2D approaches, it offers significantly improved image fidelity and structural coherence. While achieving comparable visual quality to NeRF-based methods, PGStalker delivers much faster rendering. Relative to existing 3DGS-based approaches, it further enhances perceptual quality and lip-sync accuracy, while supporting faster training and real-time inference. Overall, PGStalker achieves a strong balance between quality and efficiency, highlighting its suitability for practical applications.

\subsubsection{Cross-driven Evaluation}

As shown in Table~\ref{tab:cross-driven}, PGStalker achieves superior lip-sync accuracy under the cross-driven setting. Compared with state-of-the-art counterparts, it delivers highly competitive results with improved synchronization precision and natural facial dynamics, demonstrating strong generalization to wild audio. Notably, it consistently leads on the LMD metric while maintaining a favorable balance between LSE-C and LSE-D, indicating both temporal accuracy and perceptual stability. These results validate the robustness of our model in cross-audio scenarios and underscore its applicability to real-world talking head synthesis.

\begin{table}
	\centering 
	\caption{
    The quantitative results of the \textbf{cross-driven} setting. We extract two audio clips from the SynObama demo \cite{synobama} to drive each method and compare lip synchronization. 
    }
	\label{tab:cross-driven}
	\begin{tabular}{c@{\hskip 8pt}c@{\hskip 8pt}c@{\hskip 8pt}c@{\hskip 8pt} c@{\hskip 8pt}c@{\hskip 8pt}c}  
		\hline
		\multirow{2}{*}{\textbf{Methods}} & \multicolumn{3}{c}{\textbf{Testset A}}           & \multicolumn{3}{c}{\textbf{Testset B}}           \\  
		& LMD$\downarrow$ & LSE-C$\uparrow$ & LSE-D$\downarrow$  & LMD$\downarrow$ & LSE-C$\uparrow$    & LSE-D$\downarrow$            \\ 
		\hline
		AD-NeRF & 7.716 & 4.932 & 10.547 & 8.379 & 4.443 & 10.707          \\
		RAD-NeRF & 7.575 & \underline{6.697} & \underline{8.665} & 8.562 & \underline{6.669} & 8.620 \\
		ER-NeRF & 7.458 & \textbf{6.865} & \textbf{8.361} & 8.362 & \textbf{7.061} & \textbf{8.269} \\
		GaussianTalker & 7.994 & 6.260 & 9.523 & 8.687 & 6.618 & 8.950 \\
		TalkingGaussian & \underline{7.450} & 6.136 & 9.265 & \underline{8.323} & 6.381 & 8.637 \\
		Ours(PGStalker) & \textbf{7.426} &6.689 & 8.876 & \textbf{8.269} & 6.571 & \underline{8.551} \\
		\hline
	\end{tabular}
\end{table}

\subsection{Qualitative Evaluation}

Figure~\ref{result} presents qualitative comparisons on the Obama and May datasets under the self-driven setting. Four representative frames per method are selected to evaluate reconstruction quality and lip-sync accuracy. While 2D methods achieve reasonable lip-sync, they exhibit poor facial detail and unstable head poses. NeRF-based methods often lack natural expression rendering. In contrast, PGStalker produces more realistic results with accurate synchronization, stable poses, and fine-grained facial dynamics. Among 3DGS-based approaches, it offers the best lip-sync performance, strong visual quality, and the fastest rendering, demonstrating an effective balance between quality and efficiency.

\begin{figure}
    \centering
	\includegraphics[width=0.8\textwidth]{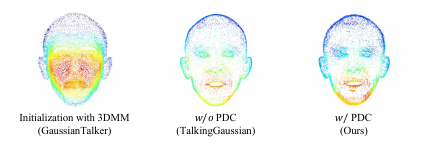}
	\caption{Visualization of Gaussian point clouds under different density control strategies. Left: GaussianTalker initialization using 3DMM exhibits uniformly high density with redundant representations. Middle: Key regions lack detail without pixel-aware density control (PDC). Right: Our method with PDC allocates higher density to dynamic areas (e.g., lips, eyes) and reduces redundancy in static regions, balancing expressiveness and efficiency.} 
	\label{mesh}
\end{figure}

\subsection{Ablation Study}

We perform ablation experiments to assess the contributions of key components in our model, including Pixel-aware Density Control (PDC), the Multimodal Gated Fusion module (MGF), and its submodules MGF-M (mouth) and MGF-F (face). As shown in Table~\ref{tab:xr}, the model consistently achieves the best performance across all metrics. Introducing PDC notably enhances image quality, reflected by improved PSNR and reduced LPIPS, demonstrating its effectiveness in allocating higher point density to critical regions. Figure~\ref{mesh} further visualizes this impact, showing clearer structural details around the lips and eyes.

\begin{table}
	\centering
	\caption{Ablation study of our contributions under the self-reconstruction setting.}
	\label{tab:xr}
	\begin{tabular}{cccccc}
		\hline
		Methods           & PSNR$\uparrow$  & LPIPS$\downarrow$   & LMD$\downarrow$   & LSE-D$\downarrow$          & LSE-C$\uparrow$          \\ \hline
		w/o PDC, MGF      & 35.09           & 0.0197              & 2.532             & 8.013                      & 7.001         \\
		w/o MGF           & 35.29           & 0.0189              & 2.538             & 7.861                      & \underline{7.247}           \\
		w/o MGF-M  & \underline{35.32} & \textbf{0.0187}          & \underline{2.474} & \textbf{7.786}             & 7.196           \\
		w/o MGF-F         & 35.27           & 0.0190              & 2.482             & 7.951                      & 7.146            \\
		All(Ours) & \textbf{35.32}     & \underline{0.0189}       & \textbf{2.469}    & \underline{7.855}          & \textbf{7.266}            \\        \hline
	\end{tabular}
\end{table}

Including the MGF module significantly boosts lip-sync performance, with notable improvements in LMD and LSE-C scores, highlighting its effectiveness in multimodal feature fusion. Further analysis shows that MGF-M and MGF-F play complementary roles: MGF-M captures fine-grained mouth motion for accurate lip movement, while MGF-F enhances overall facial consistency and expressiveness. Together, they achieve a strong balance between reconstruction quality and synchronization, validating the effectiveness of our architectural design.

\section{Conclusion}

We propose PGSTalker, a novel audio-driven talking head generation framework based on 3D Gaussian Splatting. Our approach addresses key limitations of existing methods in rendering efficiency and lip-sync accuracy. A pixel-aware density control strategy enhances point cloud representation in dynamic regions without compromising performance. In addition, a lightweight multimodal gated fusion module effectively models audio-spatial interactions, improving deformation prediction. Experiments demonstrate that PGSTalker achieves superior image quality, synchronization precision, and inference speed, highlighting its real-time capability and practical potential.

\section*{Acknowledgements}

This research was financially supported by the National Natural Science Foundation of China (Grants Nos. 62463029, 62472368, and 62303259) and the Natural Science Foundation of Tianjin (Grant No. 24JCQNJC00910).

\bibliographystyle{splncs04}
\bibliography{ref}

\begin{thebibliography}{10}
\providecommand{\url}[1]{\texttt{#1}}
\providecommand{\urlprefix}{URL }
\providecommand{\doi}[1]{https://doi.org/#1}

\bibitem{introduce1}
Chen, L., Maddox, R.K., Duan, Z., Xu, C.: Hierarchical cross-modal talking face generation with dynamic pixel-wise loss. In: {IEEE} Conference on Computer Vision and Pattern Recognition, {CVPR} 2019, Long Beach, CA, USA, June 16-20, 2019. pp. 7832--7841. Computer Vision Foundation / {IEEE} (2019)

\bibitem{cho2024gaussiantalker}
Cho, K., Lee, J., Yoon, H., Hong, Y., Ko, J., Ahn, S., Kim, S.: Gaussiantalker: Real-time high-fidelity talking head synthesis with audio-driven 3d gaussian splatting. arXiv preprint arXiv:2404.16012  (2024)

\bibitem{tvsa}
Ezzat, T., Geiger, G., Poggio, T.A.: Trainable videorealistic speech animation. {ACM} Trans. Graph.  \textbf{21}(3),  388--398 (2002)

\bibitem{guo2021ad}
Guo, Y., Chen, K., Liang, S., Liu, Y.J., Bao, H., Zhang, J.: Ad-nerf: Audio driven neural radiance fields for talking head synthesis. In: Proceedings of the IEEE/CVF international conference on computer vision. pp. 5784--5794 (2021)

\bibitem{in2}
Jamaludin, A., Chung, J.S., Zisserman, A.: You said that?: Synthesising talking faces from audio. Int. J. Comput. Vis.  \textbf{127}(11-12),  1767--1779 (2019)

\bibitem{3dgs}
Kerbl, B., Kopanas, G., Leimk{\"u}hler, T., Drettakis, G.: 3d gaussian splatting for real-time radiance field rendering. ACM Trans. Graph.  \textbf{42}(4),  139--1 (2023)

\bibitem{li2025talkinggaussian}
Li, J., Zhang, J., Bai, X., Zheng, J., Ning, X., Zhou, J., Gu, L.: Talkinggaussian: Structure-persistent 3d talking head synthesis via gaussian splatting. In: European Conference on Computer Vision. pp. 127--145. Springer (2025)

\bibitem{ernerf}
Li, J., Zhang, J., Bai, X., Zhou, J., Gu, L.: Efficient region-aware neural radiance fields for high-fidelity talking portrait synthesis. In: Proceedings of the IEEE/CVF International Conference on Computer Vision. pp. 7568--7578 (2023)

\bibitem{lsp}
Lu, Y., Chai, J., Cao, X.: Live speech portraits: real-time photorealistic talking-head animation. ACM Trans. Graph.  \textbf{40}(6) (Dec 2021)

\bibitem{dynamic3dgs}
Luiten, J., Kopanas, G., Leibe, B., Ramanan, D.: Dynamic 3d gaussians: Tracking by persistent dynamic view synthesis. In: International Conference on 3D Vision, 3DV 2024, Davos, Switzerland, March 18-21, 2024. pp. 800--809. {IEEE} (2024)

\bibitem{nerf}
Mildenhall, B., Srinivasan, P.P., Tancik, M., Barron, J.T., Ramamoorthi, R., Ng, R.: Nerf: representing scenes as neural radiance fields for view synthesis. Commun. {ACM}  \textbf{65}(1),  99--106 (2022)

\bibitem{synctalk}
Peng, Z., Hu, W., Shi, Y., Zhu, X., Zhang, X., Zhao, H., He, J., Liu, H., Fan, Z.: Synctalk: The devil is in the synchronization for talking head synthesis. In: {IEEE/CVF} Conference on Computer Vision and Pattern Recognition, {CVPR} 2024, Seattle, WA, USA, June 16-22, 2024. pp. 666--676. {IEEE} (2024)

\bibitem{wav2lip}
Prajwal, K.R., Mukhopadhyay, R., Namboodiri, V.P., Jawahar, C.: A lip sync expert is all you need for speech to lip generation in the wild. In: Proceedings of the 28th ACM International Conference on Multimedia. p. 484–492. MM '20, Association for Computing Machinery, New York, NY, USA (2020)

\bibitem{eye}
Prince, E.B., Martin, K.B., Messinger, D.S., Allen, M.: Facial action coding system. Environmental psychology \& nonverbal behavior  \textbf{1} (2015)

\bibitem{in3}
Song, L., Wu, W., Qian, C., He, R., Loy, C.C.: Everybody's talkin': Let me talk as you want. {IEEE} Trans. Inf. Forensics Secur.  \textbf{17},  585--598 (2022)

\bibitem{synobama}
Suwajanakorn, S., Seitz, S.M., Kemelmacher-Shlizerman, I.: Synthesizing obama: learning lip sync from audio. ACM Trans. Graph.  \textbf{36}(4) (Jul 2017)

\bibitem{tang2022real}
Tang, J., Wang, K., Zhou, H., Chen, X., He, D., Hu, T., Liu, J., Zeng, G., Wang, J.: Real-time neural radiance talking portrait synthesis via audio-spatial decomposition. arXiv preprint arXiv:2211.12368  (2022)

\bibitem{nvp}
Thies, J., Elgharib, M., Tewari, A., Theobalt, C., Nie{\ss}ner, M.: Neural voice puppetry: Audio-driven facial reenactment. In: Vedaldi, A., Bischof, H., Brox, T., Frahm, J. (eds.) Computer Vision - {ECCV} 2020 - 16th European Conference, Glasgow, UK, August 23-28, 2020, Proceedings, Part {XVI}. Lecture Notes in Computer Science, vol. 12361, pp. 716--731. Springer (2020)

\bibitem{mead}
Wang, K., Wu, Q., Song, L., Yang, Z., Wu, W., Qian, C., He, R., Qiao, Y., Loy, C.C.: {MEAD:} {A} large-scale audio-visual dataset for emotional talking-face generation. In: Vedaldi, A., Bischof, H., Brox, T., Frahm, J. (eds.) Computer Vision - {ECCV} 2020 - 16th European Conference, Glasgow, UK, August 23-28, 2020, Proceedings, Part {XXI}. Lecture Notes in Computer Science, vol. 12366, pp. 700--717. Springer (2020)

\bibitem{SSIM}
Wang, Z., Bovik, A., Sheikh, H., Simoncelli, E.: Image quality assessment: from error visibility to structural similarity. IEEE Transactions on Image Processing  \textbf{13}(4),  600--612 (2004)

\bibitem{x2face}
Wiles, O., Koepke, A.S., Zisserman, A.: X2face: {A} network for controlling face generation using images, audio, and pose codes. In: Ferrari, V., Hebert, M., Sminchisescu, C., Weiss, Y. (eds.) Computer Vision - {ECCV} 2018 - 15th European Conference, Munich, Germany, September 8-14, 2018, Proceedings, Part {XIII}. Lecture Notes in Computer Science, vol. 11217, pp. 690--706. Springer (2018)

\bibitem{4dgs}
Wu, G., Yi, T., Fang, J., Xie, L., Zhang, X., Wei, W., Liu, W., Tian, Q., Wang, X.: 4d gaussian splatting for real-time dynamic scene rendering. In: Proceedings of the IEEE/CVF Conference on Computer Vision and Pattern Recognition. pp. 20310--20320 (2024)

\bibitem{photo3dgs}
Yang, Z., Yang, H., Pan, Z., Zhang, L.: Real-time photorealistic dynamic scene representation and rendering with 4d gaussian splatting. In: The Twelfth International Conference on Learning Representations, {ICLR} 2024, Vienna, Austria, May 7-11, 2024. OpenReview.net (2024)

\bibitem{deformable3dgs}
Yang, Z., Gao, X., Zhou, W., Jiao, S., Zhang, Y., Jin, X.: Deformable 3d gaussians for high-fidelity monocular dynamic scene reconstruction. In: {IEEE/CVF} Conference on Computer Vision and Pattern Recognition, {CVPR} 2024, Seattle, WA, USA, June 16-22, 2024. pp. 20331--20341. {IEEE} (2024)

\bibitem{geneface++}
Ye, Z., He, J., Jiang, Z., Huang, R., Huang, J., Liu, J., Ren, Y., Yin, X., Ma, Z., Zhao, Z.: Geneface++: Generalized and stable real-time audio-driven 3d talking face generation. CoRR  \textbf{abs/2305.00787} (2023)

\bibitem{ye2023geneface}
Ye, Z., Jiang, Z., Ren, Y., Liu, J., He, J., Zhao, Z.: Geneface: Generalized and high-fidelity audio-driven 3d talking face synthesis (2023)

\bibitem{facial}
Zhang, C., Zhao, Y., Huang, Y., Zeng, M., Ni, S., Budagavi, M., Guo, X.: {FACIAL:} synthesizing dynamic talking face with implicit attribute learning. In: 2021 {IEEE/CVF} International Conference on Computer Vision, {ICCV} 2021, Montreal, QC, Canada, October 10-17, 2021. pp. 3847--3856. {IEEE} (2021)

\bibitem{LPIPS}
Zhang, R., Isola, P., Efros, A.A., Shechtman, E., Wang, O.: The unreasonable effectiveness of deep features as a perceptual metric. In: 2018 IEEE/CVF Conference on Computer Vision and Pattern Recognition. pp. 586--595 (2018)

\bibitem{sadtalker}
Zhang, W., Cun, X., Wang, X., Zhang, Y., Shen, X., Guo, Y., Shan, Y., Wang, F.: Sadtalker: Learning realistic 3d motion coefficients for stylized audio-driven single image talking face animation. In: {IEEE/CVF} Conference on Computer Vision and Pattern Recognition, {CVPR} 2023, Vancouver, BC, Canada, June 17-24, 2023. pp. 8652--8661. {IEEE} (2023)

\bibitem{pixelgs}
Zhang, Z., Hu, W., Lao, Y., He, T., Zhao, H.: Pixel-gs: Density control with pixel-aware gradient for 3d gaussian splatting. In: European Conference on Computer Vision. pp. 326--342. Springer (2024)

\bibitem{iplap}
Zhong, W., Fang, C., Cai, Y., Wei, P., Zhao, G., Lin, L., Li, G.: Identity-preserving talking face generation with landmark and appearance priors (2023)

\bibitem{zwicker2001ewa}
Zwicker, M., Pfister, H., Van~Baar, J., Gross, M.: Ewa volume splatting. In: Proceedings Visualization, 2001. VIS'01. pp. 29--538. IEEE (2001)

\end{thebibliography}
%

%
%
%
%

\end{document}